\documentstyle [12pt] {article}

\def\vev#1{\left\langle #1\right\rangle}
\begin{document}
\begin{flushright}
IUHET-252\\
UDHEP-93-03\\
BA-93-44\\
\end{flushright}
\vspace*{0.4 cm}
\begin{center}
{\bf RENORMALIZATION OF THE NEUTRINO\\
MASS OPERATOR\\ }
\vspace*{0.3 cm}
\vspace*{0.3 cm}
{\bf K.S. Babu\(^1\), C.N. Leung\(^2\), J. Pantaleone\(^3\) }\\
\vspace*{0.4 cm}
\(^1\)Bartol Research Institute\\
\(^2\)Department of Physics and Astronomy\\
University of Delaware, Newark, DE 19716 \\
\vspace*{0.4 cm}
\(^3\)Physics Department\\
Indiana University, Bloomington, IN 47405 \\
\vspace*{0.5 cm}
{\bf ABSTRACT} \\
\end{center}
A small neutrino Majorana mass can arise in the Standard
Model as an effective dimension 5 operator.
We calculate the renormalization of this operator
in the minimal Standard Model and in its two-Higgs-doublet and
supersymmetric extensions.
Renormalization from the scale of lepton number violation
(e.g., the Planck scale or a GUT scale)
to the weak scale
decreases the strength of this operator
by an order of magnitude or more if the top quark and
Higgs boson masses are large.
Neutrino mixing angles also run with momentum.
We show instances where small mixing
at a high scale becomes large at the weak scale,
and vice versa.

\newpage

{\it 1. Introduction.}
There are several indications that physical processes
do not respect the global baryon number ($B$) and lepton number ($L$)
symmetries.  For example, in the Standard Model (SM), nonperturbative
effects due to instantons break the $B+L$ symmetry \cite{instanton}.
The observed matter-antimatter asymmetry of the universe
is an explicit evidence for $B$ violation.
In addition, simple arguments suggest that quantum effects of gravity
may violate global symmetries \cite{gravity}.
It thus appears that $B$ and $L$ symmetries are
not fundamental symmetries of the full theory, but are
``accidents'' of the low
energy structure of the theory---as is the case in Grand Unified
Theories (GUT).  If lepton number is indeed not conserved,
a neutrino Majorana mass will be generated at some level.

Let \(M_X\) be the scale above which lepton number is broken.
This could be the Planck scale (if the lepton number violation
is induced only by gravity) or a GUT scale.  Assuming that the only light
fields below \(M_X\) are the ones given in the SM,\footnote{We do not
consider here the possibility of light ``exotic'' particles such as triplet
Higgs bosons, although later we shall consider
the two-Higgs-doublet and the minimal
supersymmetric extensions.} then the lowest dimension
operator which generates a Majorana mass for the left-handed neutrinos,
\(\nu_L\), is unique \cite{Weinberg}:
\begin{equation}
{\cal L}_{\nu\nu} = {1 \over 4} \kappa_{mn} \bar{l^c}^{m}_{Li} l^n_{Lj}
\Phi_k \Phi_l \epsilon_{ik} \epsilon_{jl}
+ h.c.
\label{kappa}
\end{equation}
where \(l_L\) and \(\Phi\) are the
left-handed lepton and Higgs boson doublets, respectively,
\begin{equation}
l_L^m = \left [
\begin{array}{c} \nu_L^m \\ e_L^m \end{array}
\right ]
;\ \ \ \
\Phi = \left [
\begin{array}{c}  \Phi^+ \\ \Phi^0 \end{array}
\right ] \ \ .
\end{equation}
\(\kappa_{mn}\) is symmetric under interchange of
\(m\) and \(n\), the generation indices; \(i,j,k,l\) are SU(2) indices.
When the Higgs scalar develops a vacuum expectation value,
\(\vev{\Phi^0} = v/\sqrt{2} \), the SU(2) gauge symmetry is spontaneously
broken and the neutrino Majorana mass matrix ensues:
\begin{equation}
M_\nu = - {1 \over 4} \kappa v^2 .
\end{equation}
The operator in Eq. (\ref{kappa}) has dimension 5,
so \(\kappa\) is of order \(1/M_X\).
Consequently the magnitude of the neutrino masses is
suppressed with respect to the charged fermion masses
by the factor \(v/M_X\).

Models for neutrino masses predict the coefficients, \(\kappa_{mn}\),
in terms of other parameters.
For example, if $\nu_R$'s exist, they should have a large Majorana
mass, $M_R$, from lepton number breaking and hence there will be a
contribution to $\kappa$ of
\begin{equation}
\kappa \simeq 2 Y_\nu^T M_R^{-1} Y_\nu   \ \ ,
\end{equation}
which is the popular see--saw mechanism \cite{see-saw}.
Here \(Y_\nu\) is the Yukawa coupling matrix between the
\(\l_L\)'s and \(\nu_R\)'s.  This prediction is not unique.
Models with more than one heavy scale (such as the GUT and Planck scales),
radiative models \cite{Witten,Zee},
or models with exotic heavy particles can give other predictions.
The values of \(\kappa\) depend on the short distance dynamics responsible
for lepton number breaking, while the operator structure in Eq. (\ref{kappa})
depends only on the low energy contents of the model.

Radiative corrections renormalize the neutrino mass operator,
as they do all the terms in the Lagrangian.
Studies of the running of the SM couplings have led to the
important observation that gauge coupling constants
may unify at a large scale.  The ``running'' of the neutrino mass operator
is also important for several reasons:
1) to correctly relate the scale of
lepton number violation to other physical scales,
e.g., proton decay, coupling constant unification, quantum gravity,
etc.\footnote{Solar neutrino experiments are sensitive to neutrino
masses as small as \(10^{-6}\) eV (see, e.g., \cite{KP}).
Such a small mass corresponds to lepton number breaking at a scale of order
the Planck scale,
$\kappa^{-1} \sim { v^2 \over {4 m_\nu}} \sim 10^{19}~{\rm GeV}$.}
and  2) to correctly relate the mixing angles predicted by a model
at high scales to the values at low scales where they are measured.
However, despite the fact that
the renormalization of the dimension 6 operators which contribute to
proton decay have been studied to two loops \cite{6run}, the
running of the dimension 5 term which yields a neutrino mass
has received little attention.
In this note we compute the renormalization of the neutrino mass
operator above the weak scale.\footnote{The neutrino mass does not
run significantly below the weak scale
because there the neutrino interactions with other light particles are
suppressed by the factor $q^2/M_W^2$.}
We first consider the standard model with one Higgs doublet.  Then we
shall examine its two--Higgs--doublet and supersymmetric extensions.

{\it 2. One Higgs Doublet.}
We choose to compute the renormalization of the neutrino
mass operator, Eq. (\ref{kappa}), in Landau gauge.
The topologically distinct one-loop diagrams which
give nonzero contributions are shown in Fig. 1.  Beside the
gauge interactions, there are the scalar quartic interactions,
\begin{equation}
{\cal L}_{H} = - { \lambda \over 2} ( \Phi^\dagger \Phi )^2 ,
\label{H}
\end{equation}
and the Yukawa interactions,
\begin{equation}
{\cal L}_{Y} = {\bar Q}_L \tilde{\Phi} Y_u^\dagger u_R
+ {\bar Q}_L \Phi Y_d^\dagger d_R
+ {\bar l}_L \Phi Y_e^\dagger e_R + h.c.
\label{Y}
\end{equation}
Here \(Q_L\) is the left-handed quark doublet,
\begin{equation}
Q_L = \left [
\begin{array}{c} u_L \\ d_L \end{array}
\right ]
,
\end{equation}
\(\tilde{\Phi} = i \tau_2 \Phi^*\) and \(u_R, d_R\), and \(e_R\)
are the right-handed quark and charged lepton fields.
\(Y_f\) is the Yukawa coupling matrix of fermion species $f = u, d, e$.
Generation indices have been suppressed.
With the above parametrization the physical Higgs boson mass is
\(m_{H}^2 = \lambda v^2\), where \(v = 246\) GeV.

The evolution equation for $\kappa$ is found to be
\begin{equation}
16 \pi^2 {d \kappa \over d \ln \mu}
= [ - 3 g_2^2 + 2 \lambda + 2 S ] \kappa
- {1\over 2} [\kappa (Y_e^\dagger Y_e) + (Y_e^\dagger Y_e)^T \kappa],
\label{dk1}
\end{equation}
where \(\mu\) is the renormalization scale, \(g_2\) is the SU(2)
gauge coupling constant and
\begin{equation}
S = Tr[ 3 Y_u^\dagger Y_u + 3 Y_d^\dagger Y_d + Y_e^\dagger Y_e ]
\label{S}
\end{equation}
is the contribution associated with Fig. 1c.

The presence of two Higgs fields in the neutrino mass operator
leads to qualitative and quantitative differences between
the running of \(\kappa\) and the running of the
charged fermion Yukawa couplings \cite{CEL}.
Firstly, \(\lambda\) enters Eq. (\ref{dk1}) at leading order,
so the evolution is sensitive to large Higgs boson masses.
Moreover, the evolution is twice as sensitive to a large top quark
mass.  Both of the above terms enter Eq. (\ref{dk1}) with the same
sign, so \(\kappa\) can run faster than the charged fermion
Yukawa couplings do.

Solving Eq. (\ref{dk1}), Fig. 2 plots how the magnitude of \(\kappa\)
evolves through the assumed desert above the weak scale.
We numerically evolve all relevant coupling constants
simultaneously over the entire range of \(\mu\) between $M_Z$ (the $Z$
boson mass) and $M_X$, taken here to be near the Planck scale.  (The
renormalization group equations for the SM couplings can be found in,
e.g., \cite{CEL,HLR}.)
The Yukawa couplings of the lightest two families have negligible
effects on the running of other parameters and will be ignored henceforth.
We have also ignored in Fig. 2 the small contributions from the $\tau$-lepton
Yukawa coupling.  In this approximation, all elements of $\kappa$ evolve
identically.  The parameters are chosen so that \(m_{t}\) (the top
quark mass) and \(m_{H}\)
remain small enough for perturbation theory to be valid.
We take $\alpha_s(M_Z) = 0.12$, $\alpha^{-1}(M_Z) = 128$, and
sin$^2\theta_W(M_Z) = 0.233$.
The running is significant when either \(m_{t}\) or \(m_{H}\)
is large (but especially when \(m_{t}\) is large).
Large \(m_{t}\) and \(m_{H}\) will
cause \(\kappa\) to decrease by an order of magnitude
or more in running from the Planck scale to the weak scale.

The renormalization of $\kappa$ can alter the predictions of
neutrino mass models.  For example, in SO(10) type theories
\cite{BKL,MP,BM},
if the neutrino Dirac mass matrix is identified with that of the
up--quarks and the Majorana mass matrix of $\nu_R$ is taken to be identity,
the physical neutrino masses will be given by the see--saw formula
\begin{equation}m_\nu \simeq {{m_u^2}\over {M_R}}\left[{{\kappa(M_Z)}
\over {\kappa(M_X)}}\right]\left[{{Y_u(M_X)}\over {Y_u(m_u)}}\right]^2
\end{equation}
where $m_u$ denotes the physical mass of the up--type quark and
$M_R \sim M_X$ is the Majorana mass of $\nu_R$.  Assuming an
intermediate scale $M_X \sim 10^{10}~GeV$ and taking into account the
running of $Y_u$ (but not the running of $\kappa$), the authors of
Ref. \cite{BKL} obtain
$m_{\nu_e} = 0.05 {{m_u^2}\over {M_R}},~m_{\nu_\mu} = 0.07 {{m_c^2}\over
{M_R}},~m_{\nu_\tau} = 0.18 {{m_t^2}\over {M_R}}$.
Including the running of \(\kappa\) reduces these estimates
of neutrino masses by
by an additional factor of 2 to 3 for $m_H$ near $200~GeV$.
Consequently, the intermediate
symmetry $SU(3)_c \times SU(2)_L \times SU(2)_R \times U(1)$ studied in
Ref. \cite{MP} may become compatible with the
Mikheyev-Smirnov-Wolfenstein solution \cite{MSW} of the solar neutrino
puzzle.  In contrast, large running would spoil proposals \cite{Akhmedov}
to relate the just-so solution of the solar neutrino problem
(where \(m_\nu \simeq 10^{-5}\) eV) with the Planck scale.

The running of $\kappa$ also affects model predictions of the neutrino
mixing angles.  In the approximation of two neutrino
mixing, the mixing angle is given by
\begin{equation}
{\rm tan}2\theta = {{2 \kappa_{12}} \over {\kappa_{22} - \kappa_{11}}},
\label{sine}
\end{equation}
in the basis where $Y_e$ is diagonal.
Eq. (\ref{dk1}) determines the evolution equation for the mixing
angle to be
\begin{equation}
16 \pi^2 {d {\rm sin}^2 2 \theta \over d \ln \mu}
= {\rm sin}^2 2\theta (1 - {\rm sin}^2 2\theta) (y_2^2 - y_1^2)
{{\kappa_{22} + \kappa_{11}} \over {\kappa_{22} - \kappa_{11}}},
\label{theta}
\end{equation}
where $y_1$ and $y_2$ are the Yukawa couplings of the corresponding
charged leptons.
As expected, the mixing angle does not run when the mixing is maximal
or zero.  However, significant running can occur if the magnitude of
$\kappa_{22}-\kappa_{11}$ is less than or comparable to $y_\tau^2 \sim
10^{-4}$.  (This is the case in some pseudo--Dirac \cite{W} neutrino
models.)
As a result, large mixing at $M_X$
can become small mixing at the weak scale, and vice versa.

Unlike
the elements of $\kappa$, large running of the mixing angle does not
require a heavy top quark and/or Higgs boson.
Examples are shown in Fig. 3.  We have chosen $m_t(M_Z) = 130~GeV$ and
$m_H(M_Z) = 100~GeV$.
$y_2$ is taken to be the $\tau$ Yukawa coupling, so Fig. 3 is relevant for
$e-\tau$ and $\mu-\tau$ mixing.  Curves a) and b) show that the mixing
angle can decrease drastically from its values at high energies.  This can
be achieved for neutrino masses of order $1~eV$ and
$\Delta m^2 = m_2^2 - m_1^2$ of order $10^{-5}$ to $10^{-6}~eV^2$.
Curves c) and d) illustrate the case in which small mixing at a high scale
can become large at a lower scale.  This requires
$\kappa_{22} \stackrel{_<}{_\sim} \kappa_{11}$ at the high scale,
assuming $y_2^2 > y_1^2$.  The running of the
mixing angle now depends more sensitively on the degeneracy in
$\kappa$.  This is because, according to Eq. (\ref{dk1}), $\kappa_{22}$
decreases with decreasing momentum more slowly than $\kappa_{11}$.
Consequently
they will cross at some scale, which leads to the resonance curve
shown in c).

{\it 3. Two Higgs Doublets.}
As is customary, we assume that the Higgs doublets
transform independently under the discrete symmetry
\begin{equation}
\Phi_i \rightarrow - \Phi_i
\label{symmetry}
\end{equation}
for which the most general Higgs potential is
\begin{eqnarray}
{\cal L}_{2H} & = & - { \lambda_1 \over 2} ( \Phi_1^\dagger \Phi_1 )^2
- { \lambda_2 \over 2} ( \Phi_2^\dagger \Phi_2 )^2
- \lambda_3 (\Phi_1^\dagger \Phi_1) (\Phi_2^\dagger \Phi_2)
\nonumber \\
& & - \lambda_4 (\Phi_1^\dagger \Phi_2) (\Phi_2^\dagger \Phi_1)
- [ {\lambda_5 \over 2} (\Phi_1^\dagger \Phi_2)^2 + h.c. ]
\end{eqnarray}
The discrete symmetry is to insure
that there are no flavor changing neutral
Higgs couplings in the dimension 4 terms.
With this symmetry,  each type of charged fermions can couple to
only one Higgs doublet.
We denote the doublet which couples to the charged leptons as
\(\Phi_1\).  However, there are now four ways to combine
two Higgs fields and two neutrino fields,
resulting in four operators relevant to the evolution of the
neutrino mass.  These operators fall into two classes according to
how the product $\Phi_i \Phi_j$ transform under the discrete symmetry.
Operators where $\Phi_i \Phi_j$ transform identically
will be mixed by renormalization.

The two operators in which $\Phi_i \Phi_j$ are even are
\begin{equation}
{\cal L}_{\nu\nu} = {1 \over 4}
 \kappa^{(11)}_{mn} \bar{l^c}^m_{Li} l^n_{Lj} \Phi_1^k \Phi_1^l
\epsilon_{ik} \epsilon_{jl}
+ {1 \over 4} \kappa^{(22)}_{mn} \bar{l^c}^m_{Li} l_{Lj}^n \Phi_2^k \Phi_2^l
\epsilon_{ik} \epsilon_{jl}
+ h.c. \ \ .
\end{equation}
Calculating the class of diagrams shown in Fig. 1,
we find the evolution of \(\kappa^{(11)}\) and \(\kappa^{(22)}\),
\begin{eqnarray}
16 \pi^2 {d \kappa^{(11)} \over d \ln \mu}
= [ - 3 g_2^2 + 2 \lambda_1 + 2 S^{11} ] \kappa^{(11)}
- {1\over 2} [\kappa^{(11)} (Y_e^\dagger Y_e) + (Y_e^\dagger Y_e)^T
\kappa^{(11)}]
+ 2 \lambda_5^* \kappa^{(22)}, \nonumber \\
16 \pi^2 {d \kappa^{(22)} \over d \ln \mu}
= [ - 3 g_2^2 + 2 \lambda_2 + 2 S^{22} ] \kappa^{(22)}
+ {1\over 2} [\kappa^{(22)} (Y_e^\dagger Y_e) + (Y_e^\dagger Y_e)^T
\kappa^{(22)}]
+ 2 \lambda_5 \kappa^{(11)}. \nonumber \\
\label{dk2a}
\end{eqnarray}
Here \(S^{rr}\) is the two-Higgs-doublet generalization of
Eq. (\ref{S}).  Its precise form depends on which charged fermions
\(\Phi_r\) couples to.  For instance, if $\Phi_1$ couples
to the down type quarks and the charged leptons and $\Phi_2$ couples
to the up type quarks, then
\begin{equation}
S^{11} = Tr[3 Y_d^\dagger Y_d + Y_e^\dagger Y_e],~~~~S^{22} =
Tr[3 Y_u^\dagger Y_u].
\end{equation}

The two operators in which $\Phi_i \Phi_j$ are odd are
\begin{equation}
{\cal L}_{\nu\nu} = {1 \over 2} \kappa^{(12)}_{mn} \bar{l^c}^m_{Li} l^n_{Lj}
\Phi_1^k \Phi_2^l
(\epsilon_{ik} \epsilon_{jl} - {1 \over 2} \epsilon_{ij} \epsilon_{kl} )
+ {1\over 2} \xi^{(12)}_{mn} \bar{l^c}^m_{Li} l^n_{Lj} \Phi_1^k \Phi_2^l
\epsilon_{ij} \epsilon_{kl}
+ h.c.
\end{equation}
Here \(\kappa^{(12)}_{mn}\) (and all previous \(\kappa\)'s)
are symmetric under interchange of the generation indices \(m\) and \(n\),
while \(\xi^{(12)}_{mn}\)  is antisymmetric.
Again calculating the class of diagrams shown in Fig. 1,
we find the evolution equations
\begin{eqnarray}
16 \pi^2 {d \kappa^{(12)} \over d \ln \mu}
& = & [ - 3 g_2^2 + 2 \lambda_3 + 2 \lambda_4 + S ] \kappa^{(12)}
- {1\over 2} [\kappa^{(12)} (Y_e^\dagger Y_e) + (Y_e^\dagger Y_e)^T
\kappa^{(12)}]
\nonumber \\
& & + 2 [\xi^{(12)} (Y_e^\dagger Y_e) - (Y_e^\dagger Y_e)^T \xi^{(12)}],
\nonumber \\
16 \pi^2 {d \xi^{(12)} \over d \ln \mu}
& = & [ - 9 g_2^2 + 2 \lambda_3 - 2 \lambda_4 + S ] \xi^{(12)}
+ {3\over 2} [\xi^{(12)} (Y_e^\dagger Y_e) + (Y_e^\dagger Y_e)^T
\xi^{(12)}]
\nonumber \\
& & + {3\over 2} [\kappa^{(12)} (Y_e^\dagger Y_e) -
(Y_e^\dagger Y_e)^T \kappa^{(12)}].
\label{dk2b}
\end{eqnarray}
Only the \(\kappa^{(12)}\) operator contains a neutrino mass term,
however the above equations show that the two operators
are mixed by renormalization and so \(\xi^{(12)}\) must
also be simultaneously evolved.

Different neutrino mass models yield different subsets of these
four operators below the lepton number breaking scale, \(M_X\).
For example,  the see-saw mechanism in some SO(10) models\footnote{Both
even and odd operators can be produced simultaneously if the discrete
symmetry is broken at $M_X$.} may
produce \(\kappa^{(22)}\) and \(\kappa^{(12)}\)\, while
the Zee model \cite{Zee} produces \(\kappa^{(12)}\) and \(\xi^{(12)}\).
In fact, one can easily construct a see-saw type model which leads
to any desired subset of the four operators by carefully choosing
how the different \(\nu_R\)'s transform under the discrete symmetries.
After the operators are produced, they mix under evolution according to
Eqs. (\ref{dk2a}) and (\ref{dk2b}) given above.

To illustrate the evolution of the neutrino mixing angles
in two-Higgs-doublet models, we assume that $\Phi_1$ couples to the
charged leptons and the down type quarks, while $\Phi_2$ couples to
the up type quarks.  We take the neutrino mass matrix to be
\begin{equation}
M_\nu = -{1 \over 4} v^2 [\kappa^{(11)} {\rm cos}^2\beta + \kappa^{(22)}
{\rm sin}^2\beta],
\end{equation}
as might occur in a see-saw model.  Here $v_i/\sqrt{2}$ is the vacuum
expectation value of $\Phi_i$, tan$\beta = v_2/v_1$, and $v^2 = v_1^2
+ v_2^2$.
The parameter tan$\beta$ runs according to
\begin{equation}
16 \pi^2 {{d {\rm tan}\beta} \over {d{\rm ln}\mu}} = {\rm tan}\beta
(S^{11} - S^{22}).
\end{equation}
In general, the evolution of the neutrino mixing angles
depends on the top quark and Higgs boson masses,
so large changes during running are possible.
To illustrate a particularly interesting possibility,
we have plotted in Fig. 4
the mixing angle for a two generation system as a function of momentum.
The evolution equations for the Yukawa and the quartic scalar couplings
can be found, e.g., in \cite{HLR}.
We have chosen the parameters at the weak scale to be
$y_t = 1.35$, $\lambda_1 = 0.16$, $\lambda_2 = 1.13$,
$\lambda_3 = -0.0081$, $\lambda_4 = - 0.061$, $\lambda_5 = - 0.011$,
and tan$\beta = 1$.  This corresponds to
$m_t = 168~GeV$ and a Higgs
boson spectrum of
$m_{H^{\pm}} = 47~GeV,~ m_P = 26~GeV,~ m_h = 67~GeV,~ m_H = 186~GeV$,
where $H^{\pm}$, $P$, $h$ and $H$ stand for the charged, pseudoscalar,
and the two scalar Higgs bosons.
Such a spectrum is phenomenologically acceptable
and also guarantees that the Higgs potential remains bounded
throughout the entire momentum range up to $M_X$, taken here to be
$10^{14}~GeV$.
Our choice of parameters is such that $y_t$ and $\lambda_2$ are
near their infrared fixed point values while others are not.
We then choose the \(\kappa\)'s to have,
in accordance with naive expectations,
a large hierarchy  at \(M_X\):
$\kappa^{(11)}_{11} = 0.05, \kappa^{(22)}_{12}=0.05,
\kappa^{(22)}_{22}=1.0$ (in suitable units) and all other elements to be zero.
For this parameter choice, there occurs at the weak scale
a degeneracy between the diagonal elements of the
neutrino mass matrix.
At this ``resonance'' the mixing is maximal.
The dashed line in Fig. 4 is the evolution for all the same
parameters except $\kappa^{(11)}_{11} = 0.06$.  Now the resonance
occurs at a higher momentum.
These curves illustrate that small mixing at the high scale can
become large and even maximal mixing at the weak scale.
With suitable choice of the parameters, the opposite can also happen, i.e.,
large mixing at the high scale can become small at the weak scale.

The evolution of the elements of $\kappa^{(11)}$ and $\kappa^{(22)}$ is
displayed in Fig. 5.  (All elements of $\kappa^{(11)}$ evolve identically
if the small $\tau$ Yukawa coupling is ignored, similarly for
$\kappa^{(22)}$.)  Notice that a large variation in $\kappa^{(22)}$
(by a factor of 20) is
now possible even for moderate values of $m_t$.  It is clear from
Fig. 5 that, unlike in the standard model, ``resonant mixing'' can occur in the
two-Higgs-doublet models even for nondegenerate neutrinos.
The reason is that the variation of $\kappa^{(22)}$ is
more prominent than $\kappa^{(11)}$, since $\lambda_2$
and $y_t$ are large
while $\lambda_1$ is not.  As a result, although $\kappa^{(22)}_{22} \gg
\kappa^{(11)}_{11}$  at $M_X$,
their values can become closer at lower momenta, as can be seen from
Fig. 5.  The momentum scale at which mixing angle
resonance occurs does not correspond to the momentum at which the
two $\kappa$'s cross, since the running of tan$\beta$ also affects
the evolution of the mixing angle.

{\it 4. Minimal Supersymmetric Standard Model (MSSM).}  In the MSSM,
when $R$--Parity is assumed to be an exact symmetry, lepton number
violation can arise only through the dimension 5 superpotential term
\begin{equation}
W = {1 \over 4} \kappa^{(s)}_{mn}L_i^mL_j^nH_{2k}H_{2l}\epsilon_{ik}
\epsilon_{jl}~.
\end{equation}
Here $H_2$ is the $Y=1/2$ Higgs superfield and $L$ the leptonic doublet
superfield.  Unlike in the non--SUSY two--Higgs--doublet model, the
operator in Eq. (22) does not mix with any other
operators.\footnote{There are five other dimension 5 operators in the
MSSM: $(H_1H_2)^2,~ Qu^cLe^c$, \newline $QQu^cd^c,~QQQL$ and $u^cu^cd^ce^c$.
All these carry zero $B-L$ charge and do not mix
with the operator of Eq. (22) which has $B-L = -2$.}  The evolution
equation for $\kappa^{(s)}$ is found to be
\begin{equation}
16 \pi^2 {{d \kappa^{(s)}}\over {d {\rm ln} \mu}} = \kappa^{(s)}
\left[-2g_1^2-6
g_2^2+6Tr\left(Y_u^{\dagger} Y_u\right)\right]+\kappa^{(s)} Y_e^{\dagger} Y_e
+ \left(Y_e^{\dagger}Y_e\right)^T\kappa^{(s)}~.
\label{dks}
\end{equation}

For a two--family system, the mixing angle evolution is obtained
by multiplying the right--hand side of Eq. (\ref{sine}) by $(-2)$.
In Fig. 6 we plot the running of sin$^22\theta$ as a function of $\mu$.  The
parameter choice corresponds to $y_t = y_b = y_\tau = 3$ at $M_X =
10^{16}~GeV$,
so that tan$\beta=m_t/m_b$.  The corresponding top--quark mass
is $m_t = 184~GeV$ at the weak scale.
We choose for the solid curve
$\kappa^{(s)}_{22}=1,~\kappa^{(s)}_{12}=0.1$ and $\kappa^{(s)}_{11}=0$
at $M_X$, which exhibits a
large hierarchy.  sin$^22\theta$ is seen to increase
by about a factor of 2 in running from $M_X$ to $M_Z$.  The dotted curve
corresponds to $\kappa^{(s)}_{22}=1,~\kappa^{(s)}_{12}=0.035,~
\kappa^{(s)}_{11}=0.65$ at
$M_X$.  Since $y_\tau$ in the MSSM is larger than in the SM,
the running of the mixing angle in MSSM is enhanced.  Resonant mixing can
occur even without degenerate neutrinos.

{\it 5. Conclusions.}
We have calculated the renormalization of the neutrino mass operator in the
standard model with one and two Higgs doublets, and also in the minimal
supersymmetric standard model.  The renormalization group equations are
given in Eqs. (\ref{dk1}), (\ref{dk2a}), (\ref{dk2b}) and (\ref{dks}).
These renormalization effects should be included in any neutrino mass
model.  In addition, we have solved these equations for certain
parameter choices to demonstrate some interesting features.
If the top quark mass and/or Higgs boson mass are large,
it is generally true that the neutrino
mass operators will have sizeable evolution.
Their strength can decrease by an order of magnitude
in running from the Planck scale to the weak scale.
Furthermore, running over this momentum
range can drastically change the neutrino mixing angles; possibly from
small values at high energies to large values at observable scales,
and vice versa.  Large evolution of the neutrino mixing angle requires a
near degeneracy of the neutrino masses in the standard model, but a
degeneracy is not required in models with two Higgs doublets.

{\it Note added}.  While this work was being completed we received reference
\cite{Polish} which discusses the same issues.  Our results for the
evolution equations in the supersymmetric model are in agreement.
However, there are some differences for the standard model and the
two-Higgs-doublet models, presumably due to the neglect of our
Fig. 1f in Ref. \cite{Polish}.  Our analysis is more general and reveals
the interesting possibility of resonant running of the neutrino
mixing angles.

Acknowledgements.  JP thanks the theory group at the
University of Delaware for its hospitality.
This work is supported in part by the U.S. Department of Energy
under grants No. DE-FG02-84ER40163, DE-FG02-91ER40661 and
DE-FG02-91-ER406267.

\raggedbottom

\raggedbottom
\newpage

\begin{center}
{\bf FIGURE CAPTION} \\
\end{center}
\vspace*{0.6cm}

\noindent {\bf Fig. 1} One-loop Feynman diagrams contributing to the
renormalization of $\kappa$ in Landau gauge. \\

\noindent {\bf Fig. 2} Running of \(\kappa\)
below $M_X$ = \(10^{19}\) GeV for the standard model with one Higgs doublet.
The top quark and Higgs boson masses are chosen to be, respectively, in GeV:
a) 226 and 243.2, b) 220 and 233.8, c) 180 and 183, d) 130 and 160,
e) 130 and 100. \\

\noindent {\bf Fig. 3} Illustration of the running of the neutrino mixing
angle in the standard model.  Parameter choices are:
a) $\kappa_{11}(M_X) = 1.0$ (in suitable units),
$\kappa_{22}(M_X) - \kappa_{11}(M_X) = 10^{-6}$, and $M_X = 7.88 \times
10 ^{18} GeV$; b) same as curve a) except $\kappa_{22}(M_X) - \kappa_{11}(M_X)
= 10^{-5}$; c) $\kappa_{11}(M_X) = 1.0001$ and $\kappa_{22}(M_X) = 1.0$;
d) $\kappa_{11}(M_X) = 1.0002$ and $\kappa_{22}(M_X) = 1.0$ \\

\noindent {\bf Fig. 4} Examples of the running of the neutrino mixing angle
in two-Higgs-doublet models.  Parameter choices are given in the text.\\

\noindent {\bf Fig. 5} The running of $\kappa^{(11)}$ and $\kappa^{(22)}$
corresponding to the dotted curve in Fig. 4. \\

\noindent {\bf Fig. 6} Examples of the
evolution of the neutrino mixing angle in the MSSM.
See text for the choice of parameters.

\end{document}